**Leaky-Integrate-Fire Neuron via Synthetic Antiferromagnetic Coupling and Spin-Orbit Torque**

Badsha Sekh[1], Durgesh Kumar[1], Hasibur Rahaman[1], Ramu Maddu[1], Jianpeng Chan[1], Wai Lum William Mah[1] and S.N. Piramanayagam[1,*]

[1]School of Physical and Mathematical Sciences, Nanyang Technological University, 21 Nanyang Link, 637371, Singapore

*Corresponding author: prem@ntu.edu.sg*

**Abstract**

Neuromorphic computing (NC) is a promising candidate for artificial intelligence applications. To realize NC, electronic analogues of brain components, such as synapses and neurons, must be designed. In spintronics, domain wall (DW) based magnetic tunnel junctions - which offer both synaptic and neuronal functionalities - are one of the promising candidates. An electronic neuron should exhibit leaky-integrate-fire functions similar to their biological counterparts. However, most experimental studies focused only on the integrate-and-fire functions, overlooking the leaky function. Here, we report on a domain wall neuron device that achieves integration using spin-orbit torque-induced domain wall motion and a leaky function via synthetic antiferromagnetic coupling. By fabricating Hall bar devices in a special geometry, we could achieve these two functionalities. During the leaky process, the maximum DW velocity achieved was 2500 µm/s. The proposed design utilizes materials used in STT-MRAM fabrication and is compatible with CMOS fabrication. Therefore, this neuron can be readily integrated into NC.

**Introduction**

Artificial intelligence (AI) is widely implemented across various platforms ranging from social media to astronomy[1–4]. However, the major problem associated with the present form of AI is its high-power consumption. As a result, researchers have sought inspiration from the human brain, the most intelligent device with an ultra-low power consumption. For instance, Nvidia 3090, a high-performance GPU, consumes ~ 650 W of power to perform compute-intensive tasks. In contrast, the human brain only takes 20 W of power to perform similar tasks[5]. Neuromorphic computing (NC) (or brain-inspired computing) is therefore a promising candidate for AI applications. In the human brain, neurons serve as - processors and synapses as - memory elements, and a neural network to communicate them. Similarly, electronic analogues of synapses and neurons must be designed to realize NC. Moreover, new algorithms to control the synthetic neurons and synapses must also be developed.

In the recent past, several candidates, such as phase change memory[6–9], resistive memory[10–14], ferroelectric memory[15–18], and spintronic-based devices[19–27] are being investigated for emulating the functionalities of synapses and neurons. Amongst these technologies, spintronic technology offers the virtues of low-power consumption, phenomenal miniaturization, and better endurance[28]. Within spintronics technologies, spin-torque/ spin Hall nano oscillators have been used as neurons and synapses[29–32]. As an alternative path, domain wall (DW) based magnetic tunnel junctions (MTJ) are also investigated for both synaptic and neuronfunctions[33–37]. Much focus has been dedicated to the development of synapses, while the development of neurons has been overlooked. Moreover, an electronic neuron should demonstrate both the leaky-integrate-fire and self-reset functions exhibited by their biological counterparts. However, most experimental studies lack the demonstration of leaky (and self-reset) function. It has been very difficult to achieve the leaky function experimentally, in a manner that can be adapted by the industry. For example, one proposal involves the use of an anisotropy gradient along the length of the DW device to achieve the leaky function[38,39]. However, an anisotropy gradient can be achieved only using a thickness wedge or a composition gradient created using ion-implantation, etc. These are not practical methods for CMOS-based mass-fabrication technologies. Another proposal involves the use of shape anisotropy to achieve leaky function[40]. However, the driving force for leaky function in such devices is not significantly larger than the pinning fields in these materials. As a result, there have not been many experimental demonstrations. Quite recently, experimental demonstration of leaky function has been demonstrated via synthetic antiferromagnetic coupling[41], but in this case, the integrate function was carried out through the Joule heating, which is not compatible for practical applications.

We have been investigating synthetic antiferromagnetic coupling (SAF) based neurons for achieving the leaky (and self-reset) function[42]. Although we achieved the leaky function using a stack of $(Co/Pd)_n/Ru/(Co/Pd)_n$ multilayers, it was difficult to move the DWs in those devices using spin-orbit torque (SOT)[43]. In this report, we describe our work on Co/Pt based SAF neuron device that integrates using spin-orbit torque and achieves leaky and self-reset functions via synthetic antiferromagnetic coupling. The optimized thin film stack and the methods for leaky function are entirely compatible with the CMOS fabrication techniques employed in the STT-MRAM devices and can be readily employed in NC computing.

**Micromagnetic Simulations**

To systematically understand our proposal on neuron functions through synthetic antiferromagnetic (SAF) coupling, we first performed detailed micromagnetic simulations. The simulated device is schematically presented in Figure 1 (a-c). In the proposed DW devices, we moved the DW in the soft magnetic layer. However, the magnetization state remains unchanged in the hard magnetic layer. This layer is used to facilitate the effective exchange field (from the synthetic antiferromagnetic coupling) for the leaky and self-reset processes. Therefore, we modelled the neuron devices with SAF coupling in the following manner. We considered a DW device with the dimensions of 384 nm × 64 nm × 1 nm (Figure 1 (a-c). Here, a part of the DW device (with a length of 256 nm) was considered as the region of interest (ROI) having the SAF coupling. To emulate SAF coupling, we applied an out-of-plane (OOP) magnetic field of various magnitudes along the +z direction. This means that we considered a hard magnetic layer on top of ROI with magnetization pointing along the -z-axis. The remaining 128 nm was considered to have ferromagnetic coupling with the hard magnetic layer. Therefore, we applied a constant OOP magnetic field of 1000 Oe along the -z-direction in this region. The other magnetic and geometric parameters used during the simulations are provided in methods section.

We utilized spin-orbit torque (SOT) to achieve the 'integrate' function. We studied the DW motion as a function of various relevant parameters such as SAF coupling strengths ($H_{ex}$), current density ($J$), spin Hall angle ($\theta_{SH}$), and longitudinal magnetic field ($H_x$). Moreover, we also introduced temperature ($T$ = 300 K) in our simulations to emulate near-experimental situations.

To understand the role of $H_{ex}$ on the reset (and leaky) process, we studied the forward and return DW dynamics at various $H_{ex}$ ranging from 0 Oe to 2000 Oe in the steps of 100 Oe. Here, we used $J = 7 \times 10^{11}$ A/m$^2$, $\theta_{SH}$ = 0.5, and interfacial Dzyaloshinskii–Moriya interaction (iDMI) constant = 0.5 mJ/m$^2$. The parameters mentioned above were decided based on our detailed micromagnetic simulations, which are discussed in supplementary information 1.1. All other parameters were kept the same as described in Table 1 of the methods section. As can be seen in Figure 1(a), the DW can be moved from the left end of the region of interest (ROI) to the right under the influence of the SOT in all the simulations (except for $H_{ex} \geq$ 1800 Oe). Once the DW reached the right end of the ROI, we switched off the current density and observed the DW dynamics. The pulse durations corresponding to integrate ($t_{ON}$) and reset ($t_{OFF}$) processes were optimized independently for all the studied cases and are presented in supplementary information 1.2. For $H_{ex} \leq$ 300 Oe, the coupling strength is not capable of resetting the DW position (Figure 1(d)). For 400 Oe $\leq H_{ex} \leq$ 1700 Oe, the DW resets to its initial position under the influence of SAF coupling. For $H_{ex} \geq$ 1800 Oe, the DW cannot be integrated as the SOT cannot overcome the $H_{ex}$ for the forward motion. Similar results were observed when we performed these simulations at room temperature. These results are presented in supplementary Figure S5 (a).

For deeper insight into these observations, we then plotted the forward velocity ($v_f$) as a function of $H_{ex}$ for simulations performed at 0 K as well as 300 K (Figure 1(e)). In both cases, $v_f$ decreases with the increase in $H_{ex}$. This is because the opposing torque increases as $H_{ex}$ increases. In contrast, with the increase in coupling strength, the return velocity ($v_r$) first decreases and then increases with a maximum velocity at $H_{ex}$ of 1000 Oe. With a further increase in $H_{ex}$, the return velocity decreases again. A similar trend was observed for the room temperature simulations. These results are presented in Figure 1(f). The observed results are due to at least two factors: (i) the angle of the DW surface at the start of the reset process and (ii) OOP magnetic field-driven DW motion. Therefore, we first estimated the DW surface angle for various studied cases. As can be seen in Figure 1(f), the DW surface angle takes almost a similar relation with $H_{ex}$ as the return velocity. A small discrepancy in the trend is attributed to the OOP magnetic field-driven DW motion (supplementary Figure S5(b-c)).

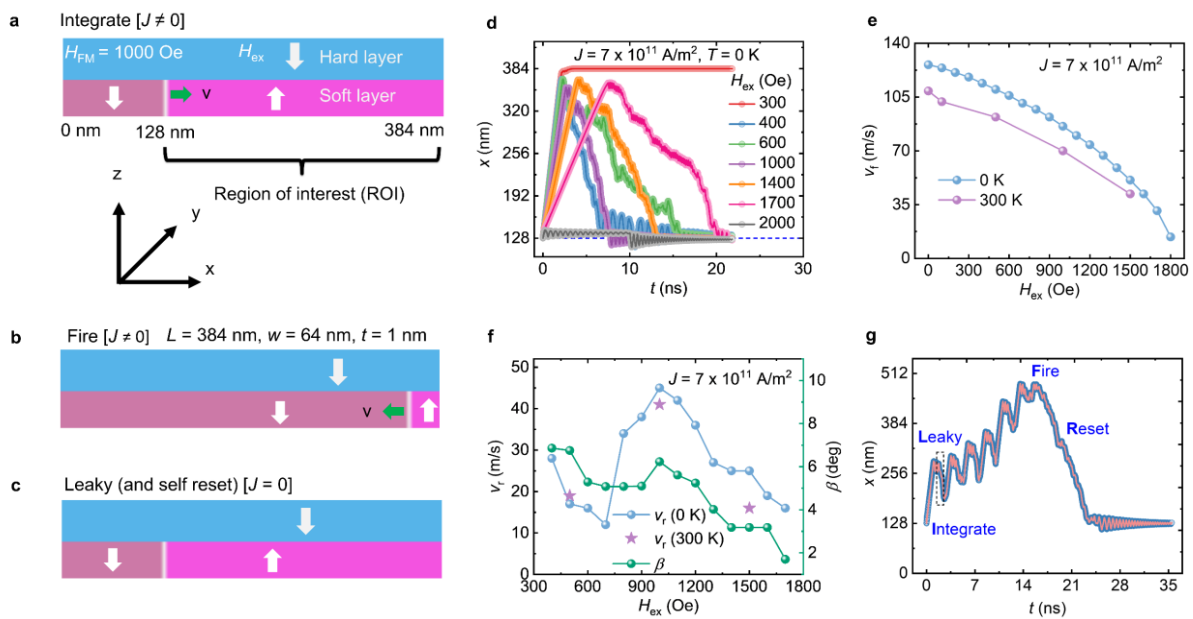

***Figure 1 Micromagnetic simulations on neuron devices with synthetic antiferromagnetic coupling:*** *(a-c) The schematic of the studied neuron device. Here, the soft magnetic layer, which carries the DW within the ROI, is synthetic antiferromagnetically coupled with the hard magnetic layer at the top. Initially, the DW can be nucleated at the start of ROI, following the energy minimization of the device.*

A non-zero current can drive the DW towards the right end of the ROI. This process is referred to as the integration process. As soon as the DW reaches the right end of the ROI, firing can be achieved. Here, we consider the read probe in the vicinity of the right end of the ROI. Once firing is realized, we switch off the current, and the DW is expected to come back to the initial position following the SAF. (d) The DW position ($x$) vs time ($t$) graph for various values of $H_{ex}$. Here, we have only presented the representative data. (e) The plot of forward velocity ($v_f$) as a function of $H_{ex}$ for simulations at 0 K as well as 300 K. (f) The plot of return velocity ($v_r$) and angle of DW surface at the start of the reset process for various studied $H_{ex}$. (g) Illustration of leaky, integrate, fire, and self-reset functions in our studied DW devices.

Finally, we performed micromagnetic simulations to study the leaky, integrate, fire, and self-reset functions in our devices. For these simulations, we increased the length of ROI to 384 nm to witness more 'integrate and leaky' events. Here, we fixed pulse durations during the integrate ($t_{ON-INT}$), leak ($t_{OFF-LEAK}$), and reset ($t_{OFF-RESET}$) at 1 ns, 1.5 ns, and 20 ns, respectively. As can be seen in Figure 1(g), we successfully achieved all the above-mentioned functions in our devices. Moreover, a total of 6 integrate and leaky events were recorded on our devices.

**Device fabrication and sample characteristics**

To study the neuron functions[44] experimentally, we deposited synthetic antiferromagnetic (SAF) stacks of Ta (1 nm)/ Pt (4.2 nm)/ [Pt (0.8 nm)/ Co (0.5nm)]$_{×2}$/ Ru ($t$ nm)/ [(Co (0.5 nm)/ Pt (0.8 nm)]$_{×4}$ (Figure 2(a)) on thermally oxidised SiO$_2$ substrates using DC magnetron-sputtering. A Pt (5 nm) layer, grown on a thin seed layer of Ta (1 nm), was used as the spin-Hall layer. The free (or soft magnetic) and hard magnetic layers of the SAF stack comprise 2 and 4 repetitions of Co/Pt multilayers, respectively. We varied Ru layer thickness to various values to achieve the optimal value of the exchange coupling field $H_{ex}$ (Supplementary Figure S7). The $H_{ex}$ vs $t_{Ru}$ exhibited three peaks. For thin values of Ru thickness (0.5 nm), the $H_{ex}$ values were very large (>10 kOe). However, for thicker values of Ru (2.55 nm), the $H_{ex}$ values were about 1040 Oe.

After the film depositions, we characterized our samples using vibrating sample magnetometry (VSM) and polar Kerr microscopy. The VSM results are presented in the supplementary Figure S8. Figure 2(c) shows a hysteresis loop obtained by Kerr microscopy for a SAF stack with Ru thickness of 2 nm. Two kinks were observed in the hysteresis loop due to the exchange field ($H_{ex}$) arising from the SAF coupling[45]. The reversal in the first quadrant at around -460 Oe arises due to the switching of the thinner multilayer. At this field, the magnetization of the thicker layer is parallel to the magnetic field and that of the thinner layer gets aligned antiparallel to the magnetic field. When the magnetic field is further increased, the magnetizations of both the layers are oriented parallel to the magnetic field direction due to Zeeman energy[46].

Subsequently, the thin films have been patterned into Hall bar devices, as shown schematically in Figure 2 (b), using optical lithography and Ar ion milling. The Hall bar comprises two regions; region A (left side of the figure) consists only soft magnetic layer, whereas region B has the SAF configuration. The underlying idea behind this design is to insert a DW at the junction of the regions A and B when an applied OOP magnetic field reduces to zero. Once the reversed domain is achieved, the DW can be moved through SOT generated from the Pt layer. The two Hall bars marked by Hall Bar 1 and Hall Bar 2 at the right end are used to observe the change in magnetization state arising due to the domain wall motion[47].

Figure 2(d) shows the anomalous Hall effect (AHE) signal from the patterned Hall bar device that shows a $H_{ex}$ value of 1150 Oe. The minor loop denoted by pink plot corresponds to the switching of bottom bilayers. The values of $H_{ex}$ measured from patterned devices were almost similar to the values measured on the thin films (Figure 2(c)). Prior to SOT-driven integration experiments, we performed the integration using an OOP magnetic field. Identical to other cases, the reset was achieved through the synthetic antiferromagnetic coupling. Later, we also performed the integration and reset experiments for multiple cycles and observed good repeatability of integration and reset functions in our neuron devices. The corresponding results have been discussed in supplementary information 2.3 in detail.

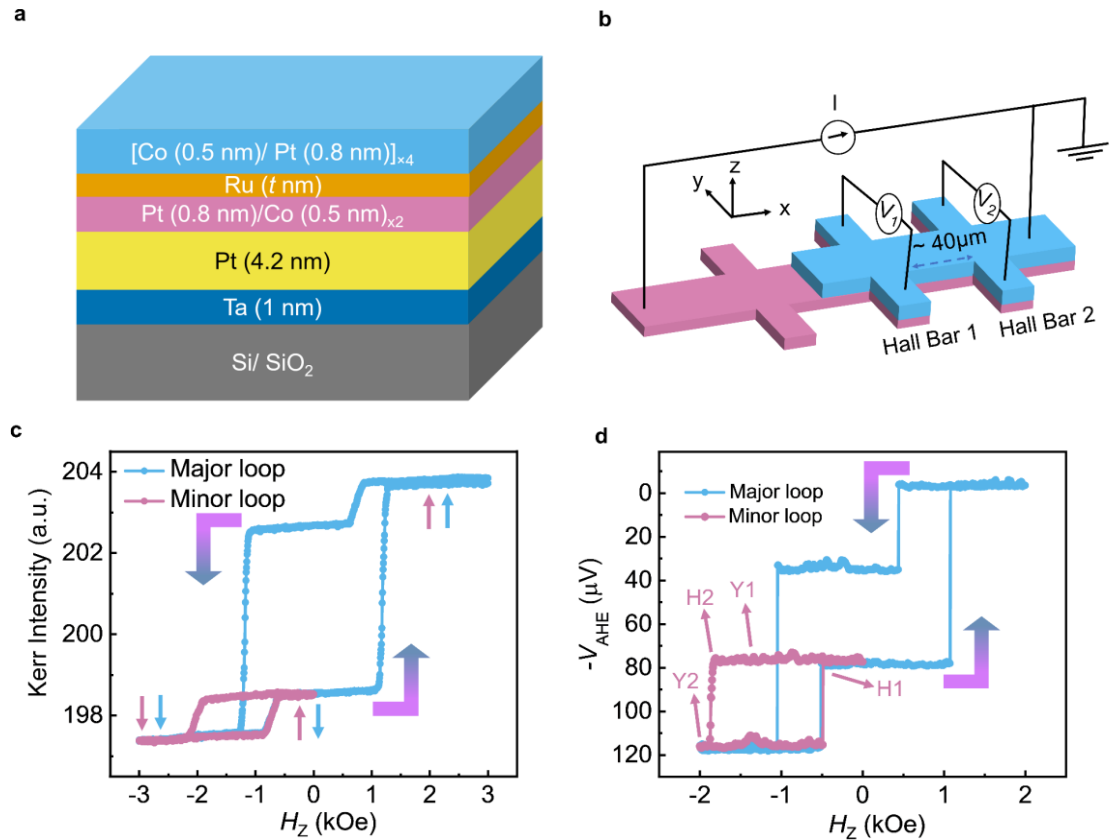

*Figure 2 Schematics of the stack, device structure and its magnetic properties.* (a) Stack structure used in this study. Ru thickness was varied from 0.5 nm to 2.4 nm. (b) The device used for AHE measurement had two regions: FM layer only and SAF region as marked by two different colors (c) Kerr hysteresis loop of the full stack with a $H_{ex}$ value of 1150 around (for Ru thickness of 2 nm). (d) The AHE measurement of the Hall bar device, where Y1 and Y1 are the AHE voltages at the ground magnetization state and saturation state respectively. H1 and H2 indicate the switching fields towards ground magnetization state and saturation state respectively.

**Integrate and Leak functions by Spin-Orbit torque and synthetic antiferromagnetic coupling**

To demonstrate the integrate function by spin-orbit torque[48], we carried out investigations on current-driven domain wall motion. The Figure 3 (a) shows an optical image of the patterned Hall bar device with the measurement configuration. A notch has been inserted at the right end of the device to stop the domain wall after full integration. Since the exchange field ($H_{ex}$) arising from SAF was so high (7100 Oe for Ru thickness of 1 nm) that SOT cannot overcome this field[49] and hence, the integration function could not be achieved. To overcome this problem, we used a stack of films with a Ru thickness of about 2 nm, wherein the $H_{ex}$ was found to be lower. Figure 2(d) shows the AHE loops, indicating that the $H_{ex}$ is 1150 Oe. From Figure 2(d), it can be inferred that an AHE voltage of about -75 μV corresponds to an antiparallel configuration of the soft and hard layers. Whereas an AHE voltage of about ~ -120 µV indicates a parallel magnetization configuration. Therefore, achieving -120 µV (from a starting point of -75 µV) means full integration in this neuron device. First, we studied the dependence of DW dynamics in the forward direction as a function of different current densities. For this, we applied $H_x$ =1000 Oe and $H_z$ = -1725 Oe (smaller than the field required to saturate the magnetizations) and current pulses with various magnitudes. For low current density values, DW motion occurs slowly with a gradual change of $V_{AHE}$ from -75 to -120 µV. However, the forward DW motion becomes faster as the current density increases. A systematic study on the effect of $H_x$ and $H_z$ revealed that the DW motion does not occur (or slow DW motion) for $H_x$ < 500 Oe (Figure 3(c)). The direction of current, $H_x$, and DW motion confirm SOT-induced domain wall motion. Additionally, we conducted a loop-shift experiment[50] to observe SOT properties of our sample stack (supplementary information 2.2). Moreover, the velocity of DW motion gets higher for higher values of $H_x$ and $H_z$. The results of the dependence of DW velocity on assisting OOP magnetic field have been shown in supplementary Figure S11 (b). In general, one does not need to apply $H_z$ to observe the domain wall motion. In this device study, the observed $H_{ex}$ is reasonably high even for 2 nm thick Ru samples, and hence a certain $H_z$ field was required to overcome the effect of $H_{ex}$ and to observe noticeable DW motion.

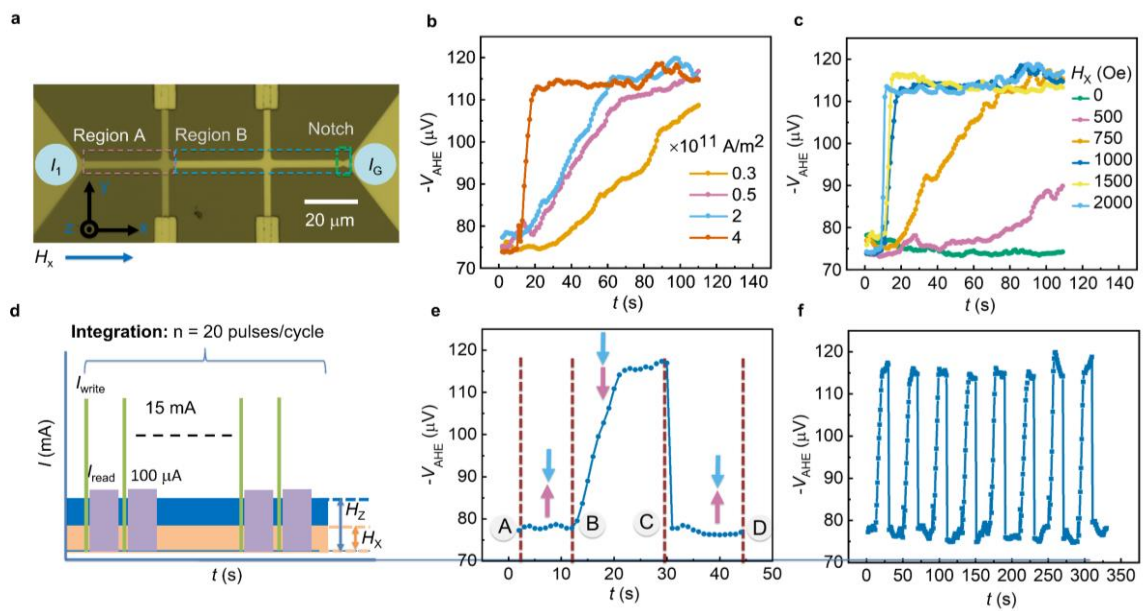

*Figure 3 Demonstration of "integrate" and "leak" functions using spin-orbit torque.* (a) Optical image of the patterned Hall bar device. Integrate function (motion of DW and the associated change in the $V_{AHE}$) observed through spin-orbit torque (b) for various current densities and (c) for various values of $H_x$. (d) The schematic of the methodology used for integration/leak function. (e) The demonstration of integration and leak functions of the studied neuron device. (f) SOT-driven 8 integration and reset cycle implies the reproducibility of the emulation of Neuron activity.

After demonstrating integrate function via SOT-induced domain wall motion, we proceeded to demonstrate the leak function. Figure 3(d) shows the schematic of the methodology used. In the presence of an $H_x$ and $H_z$ field, we applied 20 write and read current pulses as shown in the Figure 3(d). After a certain number of pulses, $V_{AHE}$ increased from -75 µV to -120 µV. Note, we first recorded the initial state with 10 pulses of only read current. When the write current pulses, $H_x$, and $H_z$ were removed, $V_{AHE}$ dropped to about -75 µV, indicating the leak (or self-reset) function. This process was repeated for 8 cycles, and in all the cycles, "integrate" and "leak" functions were observed consistently (Figure 3 (f)). Since we have used AHE to sense the magnetization, the $V_{AHE}$ increased steadily, and we can call this as a step neuron[51]. If MTJ was placed at one end, the measured voltage (or resistance) will change only when the domain wall reaches the MTJ[52]. In that case, we could observe a spike and we may call this a spike neuron.

**Return velocity investigations in samples with various $H_{ex}$**

Since $H_{ex}$ is the driving force for the leaky process[41], it is interesting to estimate the return velocity of the DW in samples with various values of $H_{ex}$. For this purpose, samples with different stack structures were prepared by varying the thickness of Ru and Co layers in the bottom bilayers. The details of the samples are presented in the supplementary information 2.4. Amongst these, we selected 4 representative samples with $H_{ex}$ values of 1500 Oe, 1220 Oe, 885 Oe, and 645 Oe and fabricated into Hall bar devices. When we attempted to estimate the return velocity by calculating the time taken for the domain wall to move to Hall Bar 1 from the notch, the motion was so fast that our setup could not be used for the estimation. Therefore, we applied an opposing OOP $H_z$ (which is slightly more than $H_2$) to slow down the return motion. As shown in Figure 2(d), $H_2$ refers to the field below which the magnetization of the soft and hard layers cannot remain parallel. In fact, $H_2$ should be equal to $H_{ex}$-$H_{c(soft)}$, where $H_{ex}$ is the exchange field and $H_{c(soft)}$ is the coercivity of the soft layer. For various values of $H_z$, the time taken for the domain walls to return (from a notch at the end) to Hall Bar 1 was calculated. Since the distance between the notch and Hall Bar 1 was kept at about 88.5 μm, the velocity can be calculated.

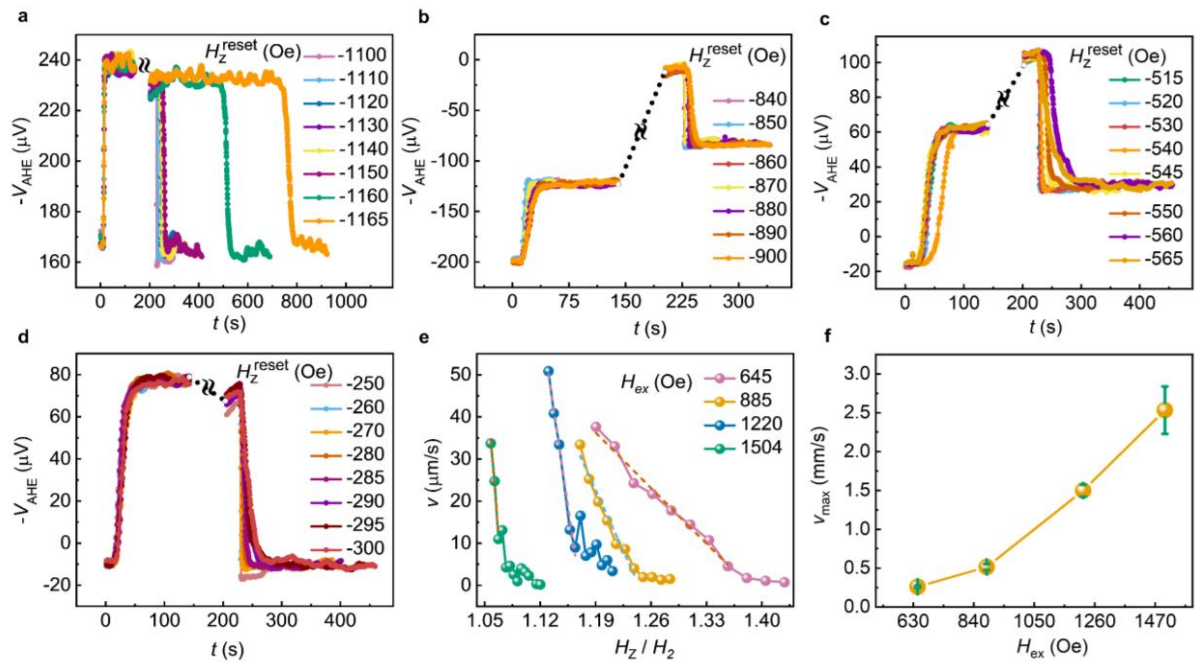

*Figure 4 Investigations of the return velocity of the domain walls, which determine the time it takes for the self-reset process.* The AHE voltages used for estimating the return velocity with (a) $H_{ex}$ = 1504 Oe, (c) $H_{ex}$ = 1220 Oe, (d) $H_{ex}$ = 885 Oe, and (e) $H_{ex}$ = 645 Oe. (e) The experimental data (solid lines) and experimental fit (dashed lines) were used for return velocity (v) measurements of samples with various $H_{ex}$. (f) The extrapolated maximum Return velocity ($v_{max}$) curve as a function of $H_{ex}$.

As can be seen in Figure 4 (a), the anomalous Hall voltage ($V_{AHE}$) vs time (t) plot provides a glimpse of a typical measurement. For a larger value of $H_z$ (-1160 Oe), the domain wall takes a very long time (about 600 s) to move from the notch to Hall Bar 1. This is because the applied $H_z$ field is larger than $H_2$ and can stabilize the parallel configuration of magnetic moments in the soft and hard magnetic layers. However, for lower values of $H_z$ field (closer to $H_2$), the domain wall motion during the reset process becomes faster. We then repeated these measurements to all the above-mentioned samples, as shown in Figure 4(b-d). For a deeper insight into the DW dynamics, we plotted DW return velocity as a function of the applied OOP magnetic field, and the results are presented in Figure 4(e). It can be noticed in Figure 4 (f), that the domain wall return velocity, in general, shows an increase with $H_{ex}$.

**Conclusions**

We have reported a DW-based artificial neuron device that emulates the integration character through SOT-driven DW motion and successfully exhibits self-reset functionality via synthetic antiferromagnetic coupling. We also have shown experimentally the systematic studies of the DW return velocity variation through modulating the interlayer exchange coupling. The return velocity increases with respect to $H_{ex}$ and reaches the maximum at the highest $H_{ex}$ value. In our micromagnetic simulations results, we observed that the forward DW velocity during the integration process decreases with $H_{ex}$ due to higher opposing torque at higher $H_{ex}$ value. On the other hand, the return velocity variation with $H_{ex}$ is related to the angle of DW during the reset process and OOP magnetic field driven DW motion. Finally, the results in our work demonstrates experimental evidence that the SAF neuron device should be integrated in neural network hardware platforms for neuromorphic computing applications.

**Methods**

**Micromagnetic Simulations**

To establish a proof-of-concept of the "leaky, integrate, fire, and self-reset" functions in neuron devices with synthetic antiferromagnetic (SAF) coupling, we first performed the micromagnetic simulations using Mumax$^3$ software[53]. In the proposed domain wall (DW) devices, we move the DW in the soft magnetic layer. However, the magnetization state remains unchanged in the hard magnetic layer. This layer is used to facilitate the effective exchange field (from the synthetic antiferromagnetic coupling) for the leaky and reset processes. Therefore, we modelled the neuron devices with SAF coupling in the following manner. We considered a DW device with the dimensions of 384 nm × 64 nm × 1 nm. Here, a part of the DW device (with a length of 256 nm; blue region in supplementary figure 1) was considered as the region of interest (ROI) having the SAF coupling. To emulate SAF coupling, we applied an out-of-plane (OOP) magnetic field of various magnitudes along the +z direction. This means that we considered a hard magnetic layer on top of ROI with magnetization pointing along the -z-axis. The remaining 128 nm (red region in supplementary Figure S1) was considered to have ferromagnetic coupling with the hard magnetic layer. Therefore, we applied a constant OOP magnetic field of 1000 Oe along the -z-direction in this region. The other magnetic and geometric parameters used during the simulations are listed in Table 1.

**Table 1:** List of geometric and magnetic properties utilized during the micromagnetic simulations[25].

| Parameter | Value |
|---|---|
| Geometrical parameters | 384 nm × 64 nm × 1 nm |
| Cell size | 1 nm × 1 nm × 1 nm |
| Exchange constant | $1.5 \times 10^{-11}$ J/m |
| Saturation magnetization | $1 \times 10^6$ A/m |
| Damping constant | 0.012 |
| DMI constant | 0.5 mJ/m$^2$ |
| Spin Hall angle | 0.5 |
| Anisotropy constant | $1 \times 10^6$ J/m$^3$ |
| Easy axis | (0, 0, 1) |
| FM coupling field | 1000 Oe |
| Temperature | 0, 300 K |

**Thin Film Deposition and Device Fabrication**

The SAF film stacks were deposited on thermally oxidized Si (100) substrates via DC/RF magnetron sputtering at room temperature. In the sputtering chamber, the substrate was rotated at 40 rpm during deposition. Prior to the sputtering, the deposition chamber has been pumped to a base pressure of 1 x 10$^{-8}$ Torr. The deposition pressure was maintained at 3 x 10$^{-3}$ Torr during sputtering. After the deposition, multilayers have been patterned into Hall bar devices by optical lithography and Ar ion milling. To nucleate a domain wall in the soft layer, some portion of the device was etched up to Ru by ion miller. The Ta(5 nm)/Cu(90 nm)/Ta(5 nm) electrodes were deposited using magnetron sputtering tool for electrical measurements. The device dimensions for field driven and SOT driven neuron (two Hall Bar device as shown the optical microscopic image in supplementary Figure S8) were 20 × 400 µm$^2$ and 5 × 150 µm$^2$ respectively. The device dimensions for SOT driven return velocity measurements (three Hall Bar device as shown in Figure 1 (b)) were 5 × 200 µm$^2$.

**Characterization**

The magnetic hysteresis loops of the multilayers were carried out by a Magvision MOKE microscopy. We utilized Lake Shore 8600 series Vibrating Sample Magnetometer (VSM) for M-H loops of our samples. The Anomalos Hall effect and magnetization switching measurements were performed using a Keithley 6221 current source and a Keithley 2182 A Nanovoltmeter, which were in-built in our MOKE system.

**Data Availability**

The data that supports the finding of this study are avialable from the corresponding author upon reasonable request.

**Acknowledgements**

The authors gratefully acknowledge the funding from the National Research Foundation (NRF), Singapore for the CRP21 grant (NRF-CRP21-2018- 0003). BS acknowledges the financial assistance from the NTU research scholarship.

**Author contributions**

Badsha Sekh: Sample preparation, device fabrication, design of experiments, device testing, analysis and writing  Durgesh Kumar: Design of experiments, Simulation, device testing, analysis and writing  Hasibur Rahaman: Simulation and analysis  Ramu Maddu: Device testing, analysis and writing  Jianpeng Chan: Device testing   Wai Lum William Mah: Conceptualization  S.N. Piramanayagam: Conceptualization of the main idea, design of experiments, analysis, interpretation and writing

**Competing interests**

The authors declare no competing interests

**Additional information**